\documentclass{cimento}

\def\fh{\hbox{$.\!\!^{\rm h}$}}

\usepackage{graphicx}
\usepackage{url}

\title{The earliest spectroscopy of the GRB 030329 
afterglow with SAO RAS 6-m telescope and early 
spectra of core-collapse supernova}

\author{V.G. Kurt\from{akc}\ETC, V.V. Sokolov\from{sao}, 
T.A. Fatkhullin\from{sao}, V.N. Komarova\from{sao},  
V.S. Lebedev\from{sao}, T.N.Sokolova\from{sao}, 
A.J. Castro-Tirado\from{spain},  
A. de Ugarte Postigo\from{spain},   
J. Gorosabel\from{spain}\from{stsci}
\atque
S. Guziy\from{ukr}\from{spain} }

\shorttitle{The earliest spectroscopy of the GRB 030329
afterglow with 6-m telescope}

\instlist{
\inst{akc} Astro Space Center of the Russian Academy of Sciences, Moscow 117810
Russia
\inst{sao} Special Astrophysical Observatory of RAS, Russia
\inst{spain} Instituto de Astrof\'{\i}sica de Andaluc\'{\i}a
(IAA-CSIC), P.O., Box 03004, E-18080 Granada, Spain
\inst{ukr} Kalinenkov Astronomical Observatory, Nikolaev State University, Nikolskaya
 24, Nikolaev, 54030, Ukraine
\inst{stsci} Space Telescope Science Institute, 3700 San Martin Drive,
Baltimore, MD 21218 USA
}
\PACSes{\PACSit{95.75.Fg}{ Spectroscopy and spectrophotometry}
\PACSit{98.70.Rz }{gamma-ray sources; gamma-ray bursts}
\PACSit{97.60.Bw}{ Supernovae}}

\begin{document}

\maketitle

\begin{abstract}
  The earliest BTA (SAO RAS 6-m telescope) spectroscopic observations 
of the GRB 030329 optical transient (OT) are presented,  
which  almost coincide  in time with the "first break" ($t\sim 0.5$ day 
after the GRB) of the OT light curve. The beginning of spectral changes 
are seen as early as $\sim 10-12$ hours after the GRB.
So, the onset of the spectral changes for $t<1$ day indicates that 
the contribution from Type Ic supernova (SN) into the OT optical flux 
can be detected earlier. The properties of early spectra of GRB 030329/SN 
2003dh  can be consistent with a shock moving into a stellar wind 
formed from the pre-SN. Such a behavior (similar to that near 
the UV shock breakout in SNe) can be explained by the existence 
of a dense matter in the immediate surroundings of massive stellar 
GRB/SN progenitor (see \cite{Young}, \cite{Imshennik}).
The urgency is emphasized of observation of early GRB/SN spectra 
for solving a question that is essential for understanding 
GRB physical mechanism: {\it Do all} long-duration gamma-ray bursts
are caused by (or physically connected to) {\it ordinary} 
core-collapse supernovae?
If clear association of normal/ordinary core-collapse SNe 
(SN Ib/c, and others SN types) and GRBs would be revealed in numbers of cases,
we may have strong observational limits for gamma-ray beaming 
and for real energetics of the GRB sources.
\end{abstract}

Spectroscopic observations of the GRB~030329 OT were performed on 29/30 March
2003  with the Multi-Pupil Fiber Spectrograph (MPFS)
(see WWW-page at \url{http://www.sao.ru/~gafan/devices/mpfs/mpfs_main.htm})
 at the 6 meter telescope (BTA) of SAO RAS starting 10.8 hours after the burst
\cite{Sokolov}. 
In order to control the absolute flux calibration,
the photometry and spectroscopy were compared.
We used $UVR_cI_c$ photometric observations of the OT carried out
with the Zeiss-1000 telescope at SAO RAS on the same epoch as the 
BTA/MPFS spectroscopy was carried out (the magnitudes and full 
$UBVR_cI_c$ light curves are reported by \cite{Gorosabel} 
and \cite{Guziy}). For the $B$ band we used the Nordic Optical 
Telescope observations \cite{Castro-Tirado}.

\begin{figure}
\hbox{
\includegraphics[width=0.45\textwidth,clip]{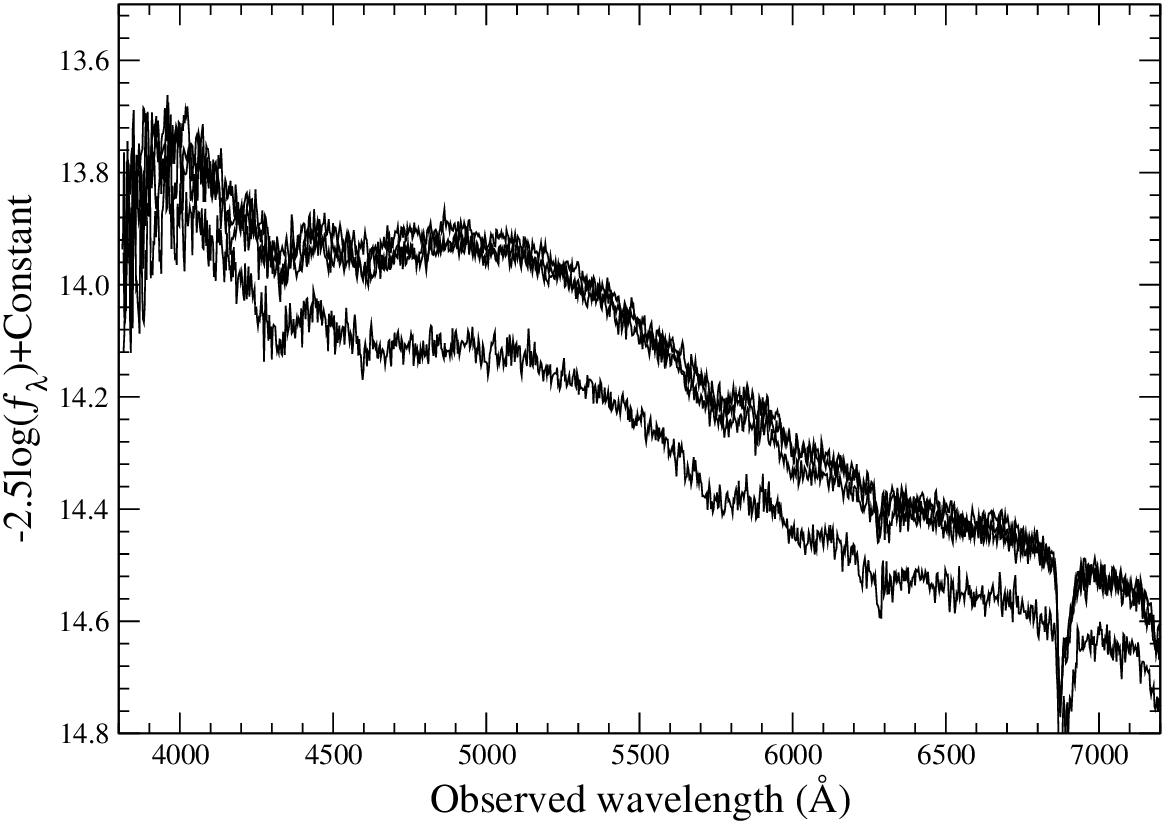}
\hfill
\hspace{1cm}
\includegraphics[width=0.45\textwidth,clip]{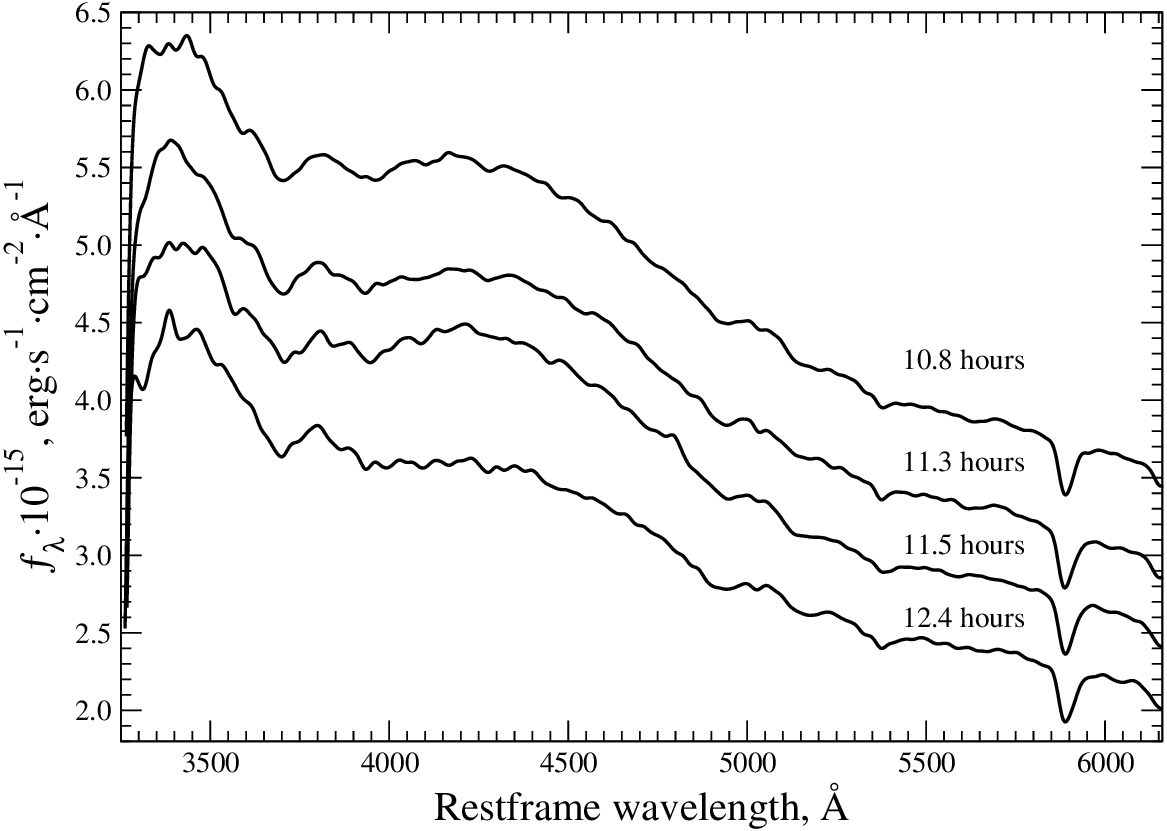}
}
\parbox[t]{0.45\textwidth}{
\caption{The four BTA/MPFS spectra of GRB~030329 OT
are presented in the order of flux decrease
for 0.45 ($10\fh8$), 0.47 ($11\fh3$), 0.48 ($11\fh5$) and 0.52 ($12\fh4$)
days after the burst, respectively.
The $10\fh8$, $11\fh3$, and $11\fh5$ spectra
correspond to almost equal $f_\lambda$.
}\label{spectra}}
\hfill
\parbox[t]{0.45\textwidth}{
\caption{The MPFS spectra (in restframe wavelengths) smoothed by a gaussian
with FWHM equal to MPFS spectral resolution (12\AA).
The smoothed spectra of GRB~030329 OT
were shifted up the scale of $f_\lambda$
relative to the last (12.4 h) spectrum.}\label{filt_spec}}
\end{figure}

Figure~\ref{spectra} presents the resulting spectra of the GRB~030329 OT.
Broad spectral features (Figures~\ref{spectra} and \ref{filt_spec})
detected in the spectra are confidently real.
As can be seen in Figure~\ref{spec-phot}, the photometry and spectroscopy
are in good agreement. The spectra showed
an unsmooth continuum with several broad spectral features
at about $4000,\ 4450,\ 5900$\AA.
From Figure~\ref{spec_evol}
one can conclude that broad spectral features
remained significant during the first three nights.
It is clearly seen from the figures that systematic deviation of V-band flux
from formal smooth power-law is due to real unsmooth OT spectra.
Also Figure~\ref{spec_evol} shows that there is some evidence of reddening
in the OT broad-band spectrum of the first three nights.
Moreover it should be noted that during about a month after the burst
the colors of the OT
are redder than during first three days,
which is in turn consistent with continuing reddening
of the broad-band spectrum. Such a behavior of
the broad-band spectrum can be explained by an increase of
a SN fraction in the GRB OT light.

We  obtained our earlier BTA/MPFS spectra of the OT
just during the most rapid variations
in the huge OT luminosity
($\sim10^{45}$\,ergs\,$s^{-1}$ at the moment $\sim 11$ h)
and physical conditions in the source,
which almost coincide in time with the first ``break''
($t\sim 0.5$ day after the GRB) of the OT light curve.
This phase is like some SNe (1993J, 1997A)
observed during the first light curves UV peaks (or during the UV
breakout phase),
specially by their similar fast luminosity variations,
spectra, and bolometric luminosities.
The bolometric luminosities in the first SNe UV peaks
can be also approximately of the same values as in the GRB OTs.

If SNe and GRBs are  indeed produced by the same astrophysical cauldron
\cite{Kawabata}, then most probably
the spectra of SN and the GRB afterglow can be mixed so closely
that it would be rather difficult to divide them in the earliest stages
of the most rapid changes in the source.
It is natural to assume that at the very beginning of the GRB/SN explosion
(in the SN rise time or in the onset time)
the contributions of {\it early} spectra of the SN
and the spectrum of the GRB afterglow can change quickly
into the common (observable) spectrum of the GRB OT.
Thus, the relative SN/OT contribution to the earliest integrated spectra might
be rapidly variable. At some moment these contributions
can become even comparable in bolometric luminosities
(as an example, see estimations of bolometric luminosity in the first maximum
of SN 1993J from \cite{Shigeyama}).
This is reinforced by the fact that the earliest spectra of SNe
are very similar to the GRB afterglow spectra in their powerful UV continuum.
And especially since (as \cite{Matheson} note for SNe Ic)
the rise time (more exactly, the beginning/onset time of the explosion)
of the majority of known SNe are not well defined
(see \cite{Norris} for more detailed discussion of this problem.)
As an illustration, we show in Fig. 5
the examples of such early spectra corresponding to the UV shock breakout in
two core collapse SNe (SN 1993J, SN 1987A),
which have the most exactly defined times of the explosion onset.

The SN 1993J was similar to a SN Ib, with a low-mass outer layer of
hydrogen (that gave the early impression of a SN II).
It can be said that its emission comes from the collision of supernova ejecta
with circumstellar gas that was released by the progenitor star
prior to the explosion. Such a  mass loss is consistent 
with the fact that the SN properties indicate that most of the stellar H
envelope is present at the time of the SN explosion.
SN 1987A is another core collapse supernova that exploded
with a (massive) H envelope. But the immediate surrounding of the 
progenitor star (the pre-SN is a blue supergiant)
was determined by the fast wind from that star.
The variety of envelopes surrounding pre-SNe is quite natural
in the evolution of a massive star
(\cite{Chevalier_a}, \cite{Chevalier_b}).

It is very important to emphasize also
that the bolometric luminosity of SN 1993J can reach the first maximum
(according to different model estimates) of order of 
$\sim10^{45}$\,ergs\,$s^{-1}$  $4 - 5$ hours after the core collapse
or $\approx$ {\it onset time} of the SN \cite{Shigeyama}.
This luminosity is approximately equal to that of GRB~030329 OT
at the moment when (at $\sim 11$ h) we obtained spectra with the BTA.

\acknowledgments
Authors are grateful to A.V. Moiseev and V.L. Afanasiev for
spectroscopic observations and help in the reduction of the spectra.
This work was supported by RFBR grant No 04-02-16300.

\begin{figure}
\hbox{
\includegraphics[width=0.4\textwidth,height=0.25\textheight,clip]{kurt_fig3.eps}
\hspace{0.1cm}
\includegraphics[width=0.55\textwidth,height=0.32\textheight,clip]{kurt_fig4.eps}
}
\parbox[t]{0.42\textwidth}{
\caption{Comparison of spectroscopy and photometry.
By horizontal bars the FWHM (full width at half maximum)
of corresponding filters are shown.}\label{spec-phot}}
\hfill
\parbox[t]{0.55\textwidth}{
\caption{Evolution of the $UBVR_cI_c$ broad-band spectra
during the first three nights. The MPFS spectra are also shown.
As in Fig.~\ref{spec-phot} by horizontal bars the FWHM 
of corresponding filters are shown.}\label{spec_evol}}
\end{figure}

\begin{figure}
\begin{center}
\hbox{
\includegraphics[width=0.45\textwidth,bb=15 15 275 208,clip]{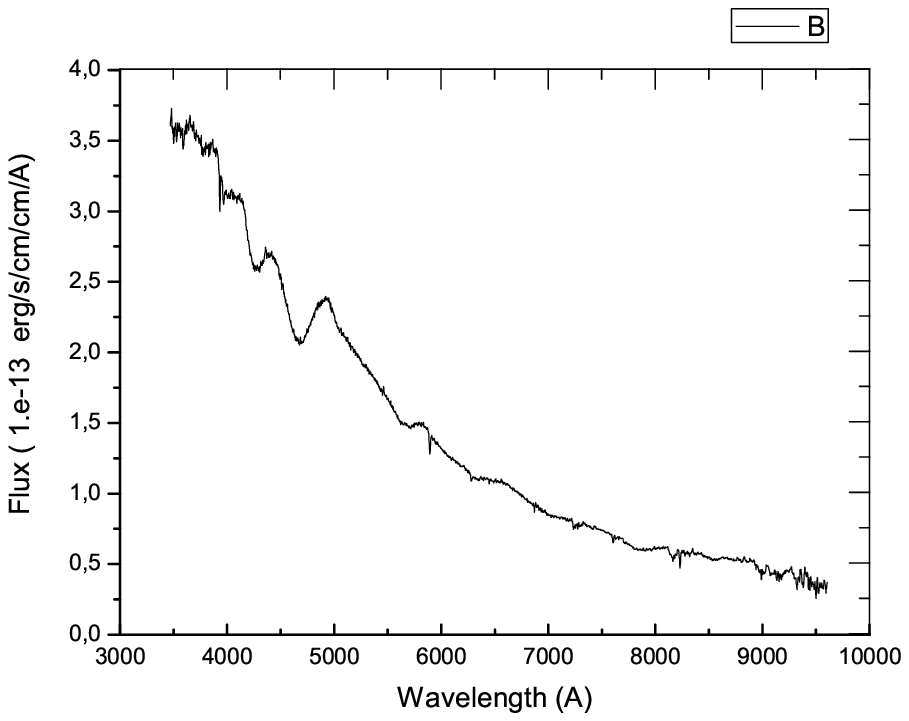}
\hspace{1cm}
\includegraphics[width=0.45\textwidth,bb=15 15 275 208,clip]{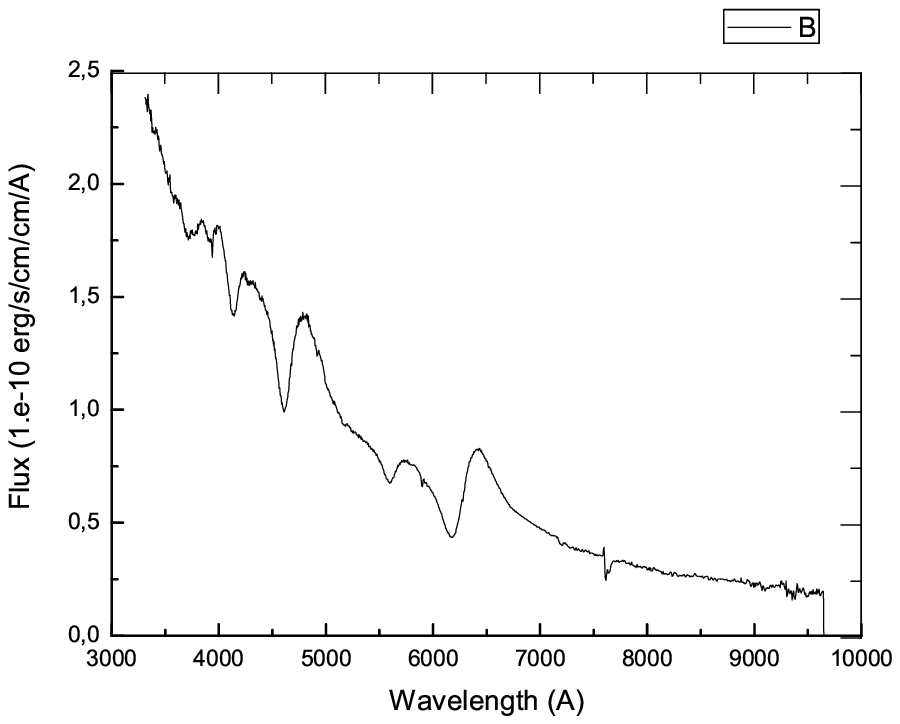}
}
\end{center}
\vspace{-1cm}
\caption{The examples of earliest optical spectra
SN 1993J and SN 1987A are given. Spectra are taken from the database
SUSPECT --- The Online Supernova Spectrum Database
http://bruford.nhn.ou.edu/~suspect/ \cite{Richardson}.}
\label{SN_spectra}
\end{figure}


\begin{thebibliography}{0}
\bibitem{Castro-Tirado} Castro-Tirado A. J., 2003, private
communication

\bibitem{Chevalier_a}
 \BY{Chevalier R.} \TITLE{The surroundings of gamma-ray bursts: constraints 
 on progenitor}, in \TITLE{Supernovae: 10 years of 1993J (IAU Col. 192)},
 edited by \NAME{Marcaide J. M. \atque Weiler K. W.} (Springer Verlag, 
 Heidelberg), 2004, p.441-451.
 
\bibitem{Chevalier_b}
 \BY{Chevalier R.} to appear in \TITLE{Young Neutron Stars and Their
  Environments (IAU Symp. 218)}, edited by \NAME{Camilo F.\atque Gaensler
  B. M.}, (ASP Conference Proceedings), 2004, astro-ph/0310730.

\bibitem{Gorosabel} \BY{Gorosabel J. et al.} \IN{A\&A}{}{2005}{submitted}.

\bibitem{Guziy} \BY{Guziy S. et al.} 2005, in preparation.

\bibitem{Imshennik} \BY{Imshennik V.S. \atque Nadyozhin D.K.} 
\IN{Uspekhi Fizicheskikh Nauk (UFN)}{vol. 5, issue 4.}{1988}{561}.

\bibitem{Kawabata}
 \BY{Kawabata K.,  Deng J.,  Wang L. et al.} \IN{ApJ}{593}{2003}{L19}.

\bibitem{Matheson}
 \BY{Matheson T.,  Garnavich P.,  Stanek K. et al.} \IN{ApJ}{599}{2003}{394}.

\bibitem{Norris}
\BY{Norris J.P. and  Bonnel  J.T.}\TITLE{How can the SN-GRB time delay be 
measured?} in \TITLE{Gamma-Ray Bursts: 30 Years of Discovery: Gamma-Ray Burst
Symposium}, edited by \NAME{Fenimore E. E. and Galassi M.}
(Melville, New York), AIP Conference Proceedings, Vol. 727, 2004, p.412-415.

\bibitem{Richardson} 
\BY{Richardson D.,  Thomas R.C.,  Casebeer D.,  
Blankenship Z.,  Ratowt S., Baron E. \atque Branch D.} 
\IN{Bull. Amer. Astron. Soc.}{34}{2002}{4}. 

\bibitem{Shigeyama}
\BY{Shigeyama T.,  Suzuki T.,  Kumagai S. et al.} \IN{ApJ}{420}{1994}{341}.

\bibitem{Sokolov} \BY{Sokolov V.V., Fatkhullin T.A., Komarova V.N.,
Kurt V.G., Lebedev V.S., Castro-Tirado A.J., Guziy S., Gorosabel J.,
de Ugarte Postigo A., Cherepaschuk A.M. \atque Postnov K.A.}, 
\IN{BSAO}{56}{2003}{5} (astro-ph/0312359)

\bibitem{Young} \BY{Young T.R., Baron E. \atque Branch D.} 
\IN{ApJ}{449}{1995}{L51}
\end{thebibliography}
\end{document}